\documentclass[10pt,a4paper,openany,twoside]{book}

\usepackage[english]{babel}
\usepackage{times,bm,url}
\usepackage[paperwidth=210mm,paperheight=285mm,textheight=240mm,textwidth=170mm,left=20mm,right=20mm, top=25mm, bottom=20mm]{geometry}
\usepackage{subfig, amsmath,amssymb,graphicx,array,multicol,fancyhdr, hhline}
\usepackage{tabularx, booktabs, multirow, slashbox, makecell}
\usepackage[perpage]{footmisc}
\usepackage[latin1]{inputenc}
\usepackage{ntheorem, titlesec,titletoc}
\usepackage{setspace}
\usepackage[section]{tocbibind}

\newcommand{\three}{2}
\renewcommand{\thesection}{\arabic{section}.}
\renewcommand{\thesubsection}{\arabic{section}.\arabic{subsection}.}
\titleformat{\chapter}{\filcenter}{}{0pt}{\bfseries\LARGE}
\titleformat{\section}{\Large\bfseries}{\thesection}{0.2em}{}
\titleformat{\subsection}{\large\bfseries}{\thesubsection}{0.2em}{}
\titlespacing*{\chapter}{0pt}{0cm}{0pt}
\titlespacing*{\section}{0pt}{3ex plus 1ex minus .2ex}{2ex  plus 1ex minus .2ex}
\titlespacing*{\subsection} {0pt}{1ex plus 1ex minus .2ex}{0ex  plus 1ex minus .2ex}
\makeatletter
\renewcommand\normalsize{%
   \@setfontsize\normalsize\@xpt\@xiipt
   \abovedisplayskip 5\p@ \@plus3\p@ \@minus5\p@
   \abovedisplayshortskip \z@ \@plus3\p@
   \belowdisplayshortskip 5\p@ \@plus3\p@ \@minus3\p@
   \belowdisplayskip\abovedisplayskip
   \let\@listi\@listI}
\makeatother
\makeatletter
\newskip\@footindent\@footindent=0.4em
\long\def\@makefntext#1{\@setpar{\@@par\@tempdima \hsize
\advance\@tempdima-\@footindent
\parshape \@ne \@footindent \@tempdima}\par
\noindent \hbox to\z@{\hss\textsuperscript{\@thefnmark}\hspace{0.2em}}#1}
\def\@makefnmark{\hbox{\textsuperscript{\@thefnmark}}}
\makeatother
\newcommand{\Ctitle}[1]{\chapter{#1}}
\newcommand{\eauthor}[1]{\vspace*{32pt}\begin{center}#1\end{center}\vspace*{-23pt}}
\newcommand{\eaddress}[1]{\begin{center}\small{#1}\end{center}\vspace*{-23pt}}
\newcommand{\email}[1]{\begin{center}\small{#1}\end{center}\vspace*{-30pt}}
\newcommand{\edate}[1]{\begin{center}\small{#1}\end{center}\vspace*{5pt}}
\newcommand{\Conauthor}[1]{\titlecontents{chapter}[0em]{}{\bfseries\contentslabel{}}{}{#1 }}

\newcommand{\eAbstract}{\noindent\textbf{\large{ABSTRACT}\vspace*{5pt}}}
\newcommand{\eKeywords}{\vspace*{10pt}\noindent\textbf{Keywords:\quad}}
\makeatletter
\def\cftitle#1{\def\@cftitle{#1}}\def\@cftitle{}
\def\ctitle#1{\def\@ctitle{#1}}\def\@ctitle{}
\def\cauthor#1{\def\@cauthor{#1}}\def\@cauthor{}
\def\publisher#1{\def\@publisher{#1}}\def\@publisher{}

\def\makecover{
    \begin{titlepage}
    \begin{center}
    \parbox[t][2cm][c]{\textwidth}{\begin{center} {\@cftitle}\end{center} }
    \parbox[t][2cm][c]{\textwidth}{\begin{center} {\Huge\@ctitle}\end{center} }
    \parbox[t][4.0cm][c]{\textwidth}{\begin{center} {\LARGE\@cauthor}\end{center} }
    \parbox[t][11.5cm][b]{\textwidth}{\begin{center} {\LARGE\@publisher} \end{center} }
    \end{center}
    \end{titlepage}
    \normalsize
   }
\makeatother
\newcommand{\headerfirstpage}{
\par\noindent\hbox to \textwidth{\textbf{\emph{Applied Mathematics, 2014, *, **}}\hfill}\vskip -1mm
\par\noindent\hbox to \textwidth{doi:******/****.***** Published Online ** 2014 \hfill}\vskip -3mm
\begin{picture}(-150,0)\put(410,10){\includegraphics[width=20mm]{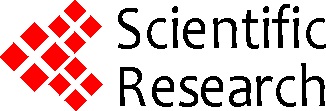}}\end{picture}}

\fancypagestyle{titlepage}{\fancyhf{}
\fancyhead[L]{\textbf{\emph{Applied Mathematics, 2014, *, **}}\\
doi:*****/****.***** Published Online ** 2014 (http://www.scirp.org/journal/am)
\begin{picture}(-150,0)
\put(60,0){\includegraphics[width=20mm]{SIGLA.jpg}}
\end{picture}}
\fancyhead[RO,LE]{}
\fancyfoot[L]{Copyright \copyright 2013 SciRes.}
\fancyfoot[R]{AM}}

\usepackage{bm}
\usepackage{hyperref}
\usepackage{float}
\usepackage{graphicx}
\usepackage{ifthen}

\theoremstyle{nonumberplain}

\begin{document}
\renewcommand\contentsname{\centerline{\Large TABLE OF CONTENTS}\vspace*{3cm}}
\renewcommand\bibname{\centerline{\normalsize REFERENCES}}
\newpage
\setcounter{page}{2019}
\setcounter{equation}{0}
\setcounter{section}{0}
\setcounter{definition}{0}
\setcounter{proposition}{0}
\setcounter{theorem}{0}
\setcounter{lemma}{0}
\setcounter{axiom}{0}
\setcounter{corollary}{0}
\setcounter{character}{0}
\setcounter{exercise}{0}


\fancypagestyle{referencespage}{
\fancyhead[LE,RO]{\thepage}                        
\fancyhead[C]{R. HURTADO\ \textit{ET}\  \textit{AL}}                         
\fancyfoot[L]{Copyright \copyright 2014 SciRes.}
\fancyfoot[R]{\textit{AM}}}                            

\ifthenelse{\three=3}{}{                                  
\Conauthor{Roger Hurtado}}                                                          
\Ctitle{Gravitational Lensing by Spherical Lenses}
\thispagestyle{titlepage}                 


\begin{spacing}{1.25}         
\eauthor{Roger Hurtado\footnotemark[1], Leonardo Casta\~neda\footnotemark[1], and Juan M. Tejeiro\footnotemark[1]}
\eaddress{\footnotemark[1]Observatorio Astron\'omico Nacional, Universidad Nacional de Colombia, Bogot\'a, Colombia}
\email{Email: rahurtadom@unal.edu.co, lcastanedac@unal.edu.co, jmtejeiros@unal.edu.co}
\end{spacing}                      
\edate{Received February 26, 2014; revised Month day, year; accepted March 10, 2014}

\eAbstract      

\noindent In this work we introduced a new proposal to study the gravitational lensing theory by spherical lenses, starting from its surface mass density $\Sigma(x)$ written in terms of a decreasing function $f$ of a dimensionless coordinate $x$ on the lens plane. The main result is the use of the function $f(x)$ to find directly the lens properties, at the same time that the lens problem is described by a first order differential equation which encodes all information about the lens. SIS and NIS profiles are used as examples to find their functions $f(x)$. Using the Poisson equation we find that the deflection angle is directly proportional to $f(x)$, and therefore the lens equation can be written in terms of the function and the parameters of the lens. The critical and caustic curves, as well as image formation and magnification generated by the lens are analyzed. As an example of this method, the properties of a lens modeled by a NFW profile are determined. Altough the puntual mass is spherically symmetric, its mass density is not continuous so that its $f(x)$ function is discussed in the Appendix A.
\eKeywords gravitational lensing: strong; dark matter


\begin{multicols}{2}
\section{Introduction}

Gravitational lensing is one of the greatest achievements of General Relativity and is one of the most useful tools of galactic astronomy, not only because the distortion of background sources carries information from the mass distribution deflecting light (called lens), but it also provides a direct test of cosmological theories \cite{blan92,yu99,brouza08,shuo12}.
\\
The deflection angle of the light, as well as the image multiplicities \cite{cohn04} and its magnifications, depend on the properties of the lens, in fact, the position and shape of the source, and the matter distribution of the lens are unknown, so you can try to resolve the problem inverting positions and shapes of the images, for expample by the Kaiser \& Squires method \cite{kaiser93}; or you can model the lens using known mass profiles, e.g. isolated mass (PM), non-singular isothermal sphere (NIS), non-singular isothermal ellipsoid (NIE), etc., depending on parameters to be adjusted so that the model reproduces the observed data \cite{knud01,halkola06}; the basis of these parametric methods rely on theoretical assumptions, which encourages to study the properties of one of the most important families of mass models: the spherically mass distribution. Due to the symmetry of this profiles, the relation between the properties of the lens-source system and its observables is reduced to a one-dimensional equation, which provides some important results from a general point of view of the theory, including image position, distortion and magnification.
\\
Of course, due to the intrinsic ellipticity of a cluster or a galaxy, it is not physically possible to model such systems using a spherical profile, however, computer simulations suggest that the dark matter halo present in these systems can be described by a mass distribution spherical\footnote{This profile is the result of the N-body simulations of collapsed structures called halos.} \cite{nfw96}, and in this sense, we shall describe our method to the NFW profile.
\\
For a basic and comprehensive reference on gravitational lensing see \cite{schn92,blan92,joach98}.
\section{Convergence and Lens Equation}
Suppose a spherically symmetric mass profile lying at a distance\footnote{All distances mentioned in this paper are angular diameter distances.} $D_{OL}$, acting as a gravitational lens on the light emitted by a source at a distance $D_{OS}$ from us, and assume that the distance between lens and source is $D_{LS}$. The mass projection on the lens plane, called surface mass density, is obtained through 
\begin{equation}\label{surfacemd}
\Sigma (x)=2\int_0^{\infty}\rho(x,z)dz,
\end{equation}
where $x$ is a dimensionless radius vector on the lens plane and the coordinate $z$ is perpendicular to it, that is to say, it is the line of sight coordinate. In this paper we suppose that
\begin{equation}\label{surfacemass}
\Sigma (x)\propto\left[f(x)+g(x)\right],
\end{equation}
where $f(x)$ and $g(x)$ are monotonically decreasing functions because the surface mass density must describe a realistic and localized lens model, and that functions are depending on the mass distribution of the lens, and defined on the interval $(0,\infty)$. Worth noting that $\Sigma(x)\geq0$, or $f(x)+g(x)\geq0$, and this can be accomplished by assuming $f(x)\geq g(x)$ and $f(x)>0$. Since $\Sigma(x)$ may be divergent at origin, we impose the condition
\begin{equation}\label{initial}
\displaystyle\lim_{x\to0}{x^2f(x)}=0.
\end{equation}
With this, the convergence is defined by
\begin{equation}\label{convergence}
\kappa(x)=\frac{1}{2C}\left[f(x)+g(x)\right],
\end{equation}
where $C$ depends on both, the distances which are functions of the cosmological model, and the physical parameters of the lens mass distribution
\begin{equation}\label{sigmacr}
\Sigma_{cr}=\frac{c^2}{4\pi G}\frac{D_{OS}}{D_{OL}D_{LS}}.
\end{equation}
Moreover, the Poisson equation relates the convergence and the deflection potential of the lens $\psi(x)$
\begin{equation}
\nabla^2\psi(x)=2\kappa(x),
\end{equation}
which, for a spherically symmetric mass distribution can be expressed as
\begin{equation}\label{poisson}
\frac{1}{x}\partial_x\psi(x)+\partial_x\partial_x\psi(x)=2\kappa(x),
\end{equation}
where $\partial_x$ denotes the derivative with respect to $x$. The Poisson equation leads to the deflection angle from
\begin{equation}
\alpha(x)=\partial_x\psi(x),
\end{equation}
thus, from equation (\ref{poisson}) can be found that
\begin{equation}
\alpha(x)=\frac{2}{x}\int_{0}^{x}x'\kappa(x')dx',
\end{equation}
or, by Eq. (\ref{convergence})
\begin{equation}\label{alphaint}
\alpha(x)=\frac{1}{C}\frac{1}{x}\int_{0}^{x}x'\left[f(x')+g(x')\right]dx'.
\end{equation}
Now, since $f(x)$ is a decreasing function $\partial_xf(x)\leq0$, and since $f(x)\geq g(x)$ we write the $g(x)$ function from $f(x)$ to which an amount $x\left|\partial_xf(x)\right|$ is subtracted, that is
\begin{equation}\label{g}
g(x)=f(x)+x\partial_xf(x),
\end{equation}
this assumption is made in order to use the fundamental theorem of calculus in the integral expression of the deflection angle, equation (\ref{alphaint}), so that
\begin{eqnarray}
\alpha(x)&=&\frac{1}{C}\frac{1}{x}\int_{0}^{x}x'\left[2f(x')+x'\partial_xf(x')\right]dx'\nonumber\\
&=&\frac{1}{C}\frac{1}{x}\int_{0}^{x}\partial_x\left[x'^2f(x')\right]dx',
\end{eqnarray}
that is
\begin{equation}\label{alpha}
\alpha(x)=\frac{1}{C}xf(x).
\end{equation}
The anterior result shows that for a spherically symmetric mass profile, whose surface mass density can be written in the form of Equation (\ref{surfacemass}), the deflection angle is proportiional to the function $f(x)$.
\\
The lens equation, which relates the image and source positions, $x$ and $y$ respectively, for a spherically symmetric situation, is a scalar and takes the one dimensional form,
\begin{equation}
y(x)=x-\alpha(x),
\end{equation}
which can be written in terms of the $f(x)$ function as
\begin{equation}\label{lensequation}
y(x)=x|1-\frac{1}{C}f(x)|.
\end{equation}
Joining the results given above, the $f(x)$ function satisfies the following equation

\begin{equation}\label{differentialf}
\boxed{
x\frac{df(x)}{dx}+2f(x)-2C\kappa(x)=0,}
\end{equation}

which comes from inserting Equation (\ref{g}) in Equation (\ref{convergence}), according to the initial condition (\ref{initial}). Thus, the problem is reduced to solve the first-order ordinary differential equation Eq. (\ref{differentialf}) for $f(x)$.
\subsection{$f(x)$ for SIS and NIS profiles}\label{SISNIS}
A spherical model widely used in the gravitational lensing theory is the singular isothermal sphere (SIS) \cite{schn92}, whose convergence is given by
\begin{equation}\label{SISconvergence}
\kappa_{S}(x)=\frac{2\pi\sigma^2}{c^2}\frac{D_{LS}}{D_{OS}}\frac{1}{x},
\end{equation}
where $\sigma_v$ is the one-dimensional velocity dispersion. With the Equation (\ref{SISconvergence}) plugged into Equation (\ref{differentialf}) and the Equation (\ref{initial}), one obtains the $f_{S}(x)$ function for a SIS
\begin{equation}
x\frac{df_{S}(x)}{dx}+2f_{S}(x)-\frac{1}{x}=0,
\end{equation}
\begin{equation}\label{functionfsis}
f_{S}(x)=\frac{1}{x},
\end{equation}
with
\begin{equation}\label{cis}
C_{S}=\frac{c^2}{4\pi\sigma^2}\frac{D_{OS}}{D_{LS}}.
\end{equation}
The $g_{S}(x)$ function is then for a SIS
\begin{equation}\label{functiongsis}
g_{S}(x)=0.
\end{equation}
To find the deflection angle, make the product $x$ with $f_{S}(x)/C_{S}$
\begin{equation}
\alpha_{S}(x)=C_{S}^{-1}.
\end{equation}
One generalization of the SIS model is frequently used with a finite core $x_0$, that is the non-singular isothermal sphere (NIS), which is more realistic for modeling galaxies. In this case, the convergence is given by
\begin{equation}
\kappa_{N}(x)=\frac{2\pi\sigma^2}{c^2}\frac{D_{LS}}{D_{OS}}\frac{2x_0^2+x^2}{\left(x_0^2+x^2\right)^{3/2}},
\end{equation}
Through a procces similar to the SIS, we can found $f_{N}(x)$, $g_{N}(x)$ and $\alpha_{N}(x)$ for the NIS profile
\begin{equation}\label{functionfnis}
f_{N}(x)=\frac{1}{(x_0^2+x^2)^{1/2}},
\end{equation}
\begin{equation}\label{functiongnis}
g_{N}(x)=\frac{x_0^2}{(x_0^2+x^2)^{3/2}},
\end{equation}
and
\begin{equation}
\alpha_{N}(x)=C_{S}^{-1}\frac{x}{(x_0^2+x^2)^{1/2}},
\end{equation}
where $C_{S}$ is given by Equation (\ref{cis}).

\section{Magnification and Shear}
Since gravitational lensing conserves the brightness of a source, the magnification of an image is defined as the ratio between the solid angles of the image and the source. Namely
\begin{equation}
\mu(x)=\left[\frac{y}{x}\partial_xy(x)\right]^{-1},
\end{equation}
from Eq. (\ref{g}) and Eq. (\ref{lensequation}), this is
\begin{equation}\label{mu}
\mu(x)=\frac{C^2}{|C-f(x)||C-g(x)|}.
\end{equation}
Eq. (\ref{mu}) implies that the magnification has two singularities in $f(x)=C$ and $g(x)=C$ and therefore its curve has two asymptotes at these points. In the next section we will see that those points in the lens plane for $f(x)=C$ and $g(x)=C$ are the critical points.
\\
Noting that the magnification Eq. (\ref{mu}) can be written in terms of the convergence $\kappa(x)$ and shear $\gamma(x)$, which measures the distortion of images,
\begin{equation}
\mu(x)=\left(\left[1-\kappa(x)\right]^2-\gamma(x)^2\right)^{-1},
\end{equation}
whereby
\begin{equation}
\gamma(x)^2=\left[1-\kappa(x)\right]^2-\left[1-\frac{1}{C}f(x)\right]\left[1-\frac{1}{C}g(x)\right],
\end{equation}
and from Eq. (\ref{convergence}), the shear is
\begin{equation}\label{gamma}
\gamma(x)=\frac{1}{2C}\left[f(x)-g(x)\right].
\end{equation}
This expression allows to calculate $\gamma(x)$ in a straightforward way. For example, returning to the models shown in \S \ref{SISNIS}, through the Eq. (\ref{functionfsis}) and Eq. (\ref{functiongsis}), the shear generated by a SIS profile is
\begin{equation}
\gamma_S(x)=\frac{1}{2C}\frac{1}{x}
\end{equation}
and that generated by a NIS profile
\begin{equation}
\gamma_N(x)=\frac{1}{2C}\frac{x^2}{\left(x_0^2+x^2\right)^{3/2}}
\end{equation}
where we made use of Eq. (\ref{functionfnis}) and Eq. (\ref{functiongnis}).
\\
Now, recognizing that 
\begin{equation}
f(x)=C\frac{1}{x}\partial_x\psi(x),
\end{equation}
and
\begin{equation}
g(x)=C\partial_x\partial_x\psi(x),
\end{equation}
the shear can be written in terms of the deflection potential of a mass distribution with spherical symmetry, as
\begin{equation}
\frac{1}{x}\partial_x\psi(x)-\partial_x\partial_x\psi(x)=2\gamma(x).
\end{equation}
Here, the definition of the $f(x)$ function shows again its usefulness, since the shear can be found in terms of the deflection potential without have recourse to the partial derivatives of it.
\section{Critical and Caustics curves}
The critical curves are those points $x$ in the lens plane where the lens equation can not be inverted, or equivalently, those points where the magnification is infinite, which satisfy
\begin{equation}
\left[1-\kappa(x)\right]^2-\gamma(x)^2=0,
\end{equation}
or
\begin{eqnarray}
\kappa(x)+\gamma(x) & = & 1 , \\
\kappa(x)-\gamma(x) & = & 1 ,
\end{eqnarray}
but, from Eq. (\ref{convergence}) and Eq. (\ref{gamma})
\begin{equation}
\kappa(x)+\gamma(x)=\frac{1}{C}f(x),
\end{equation}
and
\begin{equation}
\kappa(x)-\gamma(x)=\frac{1}{C}g(x),
\end{equation}
Thus, the critical curves are the level contours of the $f(x)$ and $g(x)$ functions, and are found when
\begin{equation}\label{critical}
f(x_{c1})=C\qquad\mbox{or}\qquad g(x_{c2})=C.
\end{equation}
The equations (\ref{critical}) are not associated forming a system, thus, given $C$ it is possible found two solutions $x_{c1}$ and $x_{c2}$ if $g(x)>0$, or $\partial_x\left[xf(x)\right]>0$, meaning that $xf(x)$ is increasing, and, if $g(x)\neq a$, so that $\partial_x\left[xf(x)\right]\neq0$ and therefore if $f(x)\neq a+b/x$, with $a$ and $b$ two arbitrary constants. In this way, the only condition for forming two critital circles is that $xf(x)$ is incresing, otherwise the lens produces only a single critical curve associated to $f(x)$, or any critical curve if $f(x)<C$. At the same time, the caustics curves are the corresponding locations in the source plane of the critical curves through the lens equation, and if we assume that the lens produces two critical curves, that is,
\begin{eqnarray}
y(x_{c1}) & = & 0 , \\
y(x_{c2}) & = & -\frac{1}{C}x_{c2}^2\partial_xf(x_{c2}) ,
\end{eqnarray}
with $\partial_xf(x)\leq0$ on $x\in(0,\infty)$. Thus, caustic curves will be a point and a circle concentric with the lens.
\section{Image formation}\label{forimage}
\begin{figure}[H]
	\includegraphics[width=84mm]{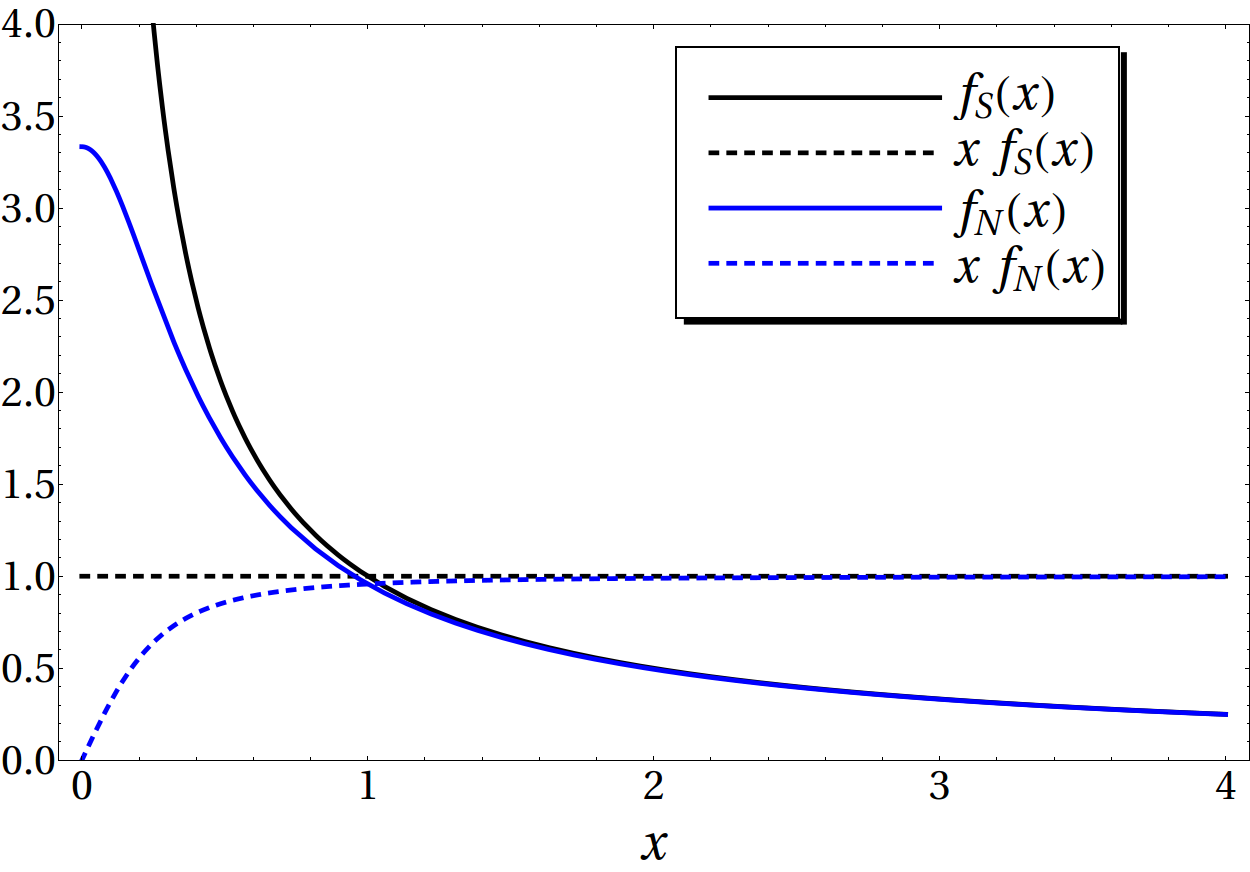}
\caption{Function $f(x)$ for two lens models, the singular isothermal sphere and the Non-singular isothermal sphere (NIS). In the case of SIS $xf(x)$ is constant, while $f(x)$ is decreasing. In the NIS model $xf(x)$ is increased, unlike $f(x)$.}
\label{figu1}
\end{figure}

In general, the image multiplicity depends on the source position with respect to the caustic circle, changing in two as the source crosses through it. Moreover caustics depends on the critical curves and on the increase or decrease of $xf(x)$ as seen in the previous section. The Fig. (\ref{figu1}) shows the two basic sketches for the function $f(x)$ for the two lens models mentioned in section \S \ref{SISNIS}: the SIS and the NIS profiles, where we can see that although $f(x)$ is decreasing, the product $xf(x)$ can be increased or constant, but this depends on the lens model as follows:
\begin{itemize}
\item If $f(x)>C$ in $(0,x_{c1})$ and $xf(x)$ is increasing, $x_{c2}$ does exist and the maximum number of images are three.
\item If $f(x)<C$ in $(0,x_{c1})$ and $xf(x)$ is decreasing, $x_{c2}$ does not exist and the total number of images are two.
\item If $f(x)<C$ for $x>0$, then $xf(x)$ is increasing and there is only one image.
\end{itemize}
In the case where the lens produces three images the source is inside the caustic circle, that is $y<y(x_{c2})$ and let us call those images, $\alpha$, $\beta$ and $\gamma$, which together obey
\begin{equation}
0<x_{\alpha}<x_{c2}<x_{\beta}<x_{c1}<x_{\gamma},
\end{equation}
or
\begin{equation}
x_j<x_{c1},
\end{equation}
with $j=\alpha,\beta$; and since $f(x)$ is decreasing in $x\in(0,\infty)$,
\begin{equation}\label{fcimage}
f(x_j)>f(x_{c1})=C.
\end{equation}
Now suppose the source located in the first quadrant of a cartesian coordinate system in whose center lies the lens; namely $y_i>0$ with $i=1,2$. The lens mapping (\ref{lensequation}), is, by components
\begin{equation}
y_i=x_i\left[1-\frac{1}{C}f(x)\right],
\end{equation}
through Equation (\ref{fcimage})
\begin{equation}
1-\frac{1}{C}f(x_j)<0,\quad\mbox{con}\quad j=\alpha,\beta.
\end{equation}
This implies
\begin{equation}
x_{ji}<0,
\end{equation}
therefore, the images $\alpha$ and $\beta$ will be in the third quadrant. And since the angle of $\bm y=(y_1,y_2)$ is,
\begin{equation}
\frac{y_2}{y_1}=\frac{x_{j2}}{x_{j1}},
\end{equation}
the angle of images will be
\begin{equation}
\arctan\left(\frac{x_{j2}}{x_{j1}}\right)=\pi+\arctan\left(\frac{y_2}{y_1}\right),
\end{equation}
that is, the images $\alpha$ and $\beta$ lies on the same line connecting the source and lens, but are diametrically opposed to the latter one.
\\
Meanwhile, the third image $\gamma$ satisfies $x_\gamma>x_{c1}$, as
\begin{equation}
1-\frac{1}{C}f(x_{\gamma})>0,
\end{equation}
and since $y_i>0$, it is found $x_{\gamma i}>0$, with $i=1,2$. This implies that the angle of $\bm x_{\gamma}$ is equal to that of the source. The third image also lies in the same line between lens and source.
\\
If the lens produces only two images, they are diametrically opposed lying on the line lens-source. And, if the lens produces only one image, it will be located at the same angle of the source.
\section{Applying theory to the NFW profile}
The lensing effects of the NFW profile have been widely studied \cite{wright2000,golse2002,nari12}. The NFW profile describes the distribution of a dark matter halo. The dark matter halo is useful to calculate the function $f(x)$ in this model.\\
Suppose a gravitational lens modeled by a NFW profile \cite{nfw96} with a mass density given by
\begin{equation}\label{NFW}
\rho(r)=\frac{\rho_0}{(r/r_s)(1+r/r_s)^2},
\end{equation}
where the so called scale radius $r_s$ and $\rho_0$ are parameters of the halo.
\subsection{NFW Convergence and Lens Equation}
The NFW mass density Eq. (\ref{NFW}) expressed in terms of $r=(\xi^2+z^2)^{1/2}$, where $\xi$ is a radius vector on the lens plane, and the Equation (\ref{surfacemd}) leads to the convergence through $\Sigma_{cr}$, Eq. (\ref{sigmacr}), to obtain

\begin{equation}\label{convergenceNFW}
\kappa(x)=-\frac{1}{2C(1-x^2)}\left(1-\frac{2}{\left(1-x^2\right)^{1/2}}\mbox{ArcTanh}\left[\frac{\left(1-x\right)^{1/2}}{\left(1+x\right)^{1/2}}\right]\right),
\end{equation}
this expression is according with the results found in \cite{bartelmann} and \cite{wright2000}. Here we have defined $x=\xi/r_s$ and
\begin{equation}\label{cconstant}
C=\frac{\Sigma_{cr}}{4\rho_0r_s}.
\end{equation}
The Eq. (\ref{differentialf}) and Eq. (\ref{convergenceNFW}) leads to the differential equation
\begin{eqnarray}
\frac{df(x)}{dx}+\frac{2}{x}f(x)+\frac{1}{x(1-x^2)}\left(1-\frac{2}{\left(1-x^2\right)^{1/2}}\times\right.\nonumber\\
\left.\mbox{ArcTanh}\left[\frac{\left(1-x\right)^{1/2}}{\left(1+x\right)^{1/2}}\right]\right)=0,\quad
\end{eqnarray}

finding that
\begin{equation}\label{functionf}
f(x)=\frac{1}{x^2}\left(\ln\left(\frac{x}{2}\right)+\frac{2}{\left(1-x^2\right)^{1/2}}\mbox{ArcTanh}\left[\frac{(1-x)^{1/2}}{(1+x)^{1/2}}\right]\right),
\end{equation}
where we use the Eq. (\ref{initial}). Therefore
\begin{eqnarray}\label{functiong}
g(x)&=&-\frac{1}{x^2}\left(\ln\left(\frac{x}{2}\right)+\frac{x^2}{1-x^2}+\right.\nonumber\\
&&\qquad\left.\frac{2\left(1-2x^2\right)}{\left(1-x^2\right)^{3/2}} \mbox{ArcTanh}\left[\frac{(1-x )^{1/2}}{(1+x )^{1/2}}\right]\right).\qquad
\end{eqnarray}

\begin{figure}[H]
	\includegraphics[width=84mm]{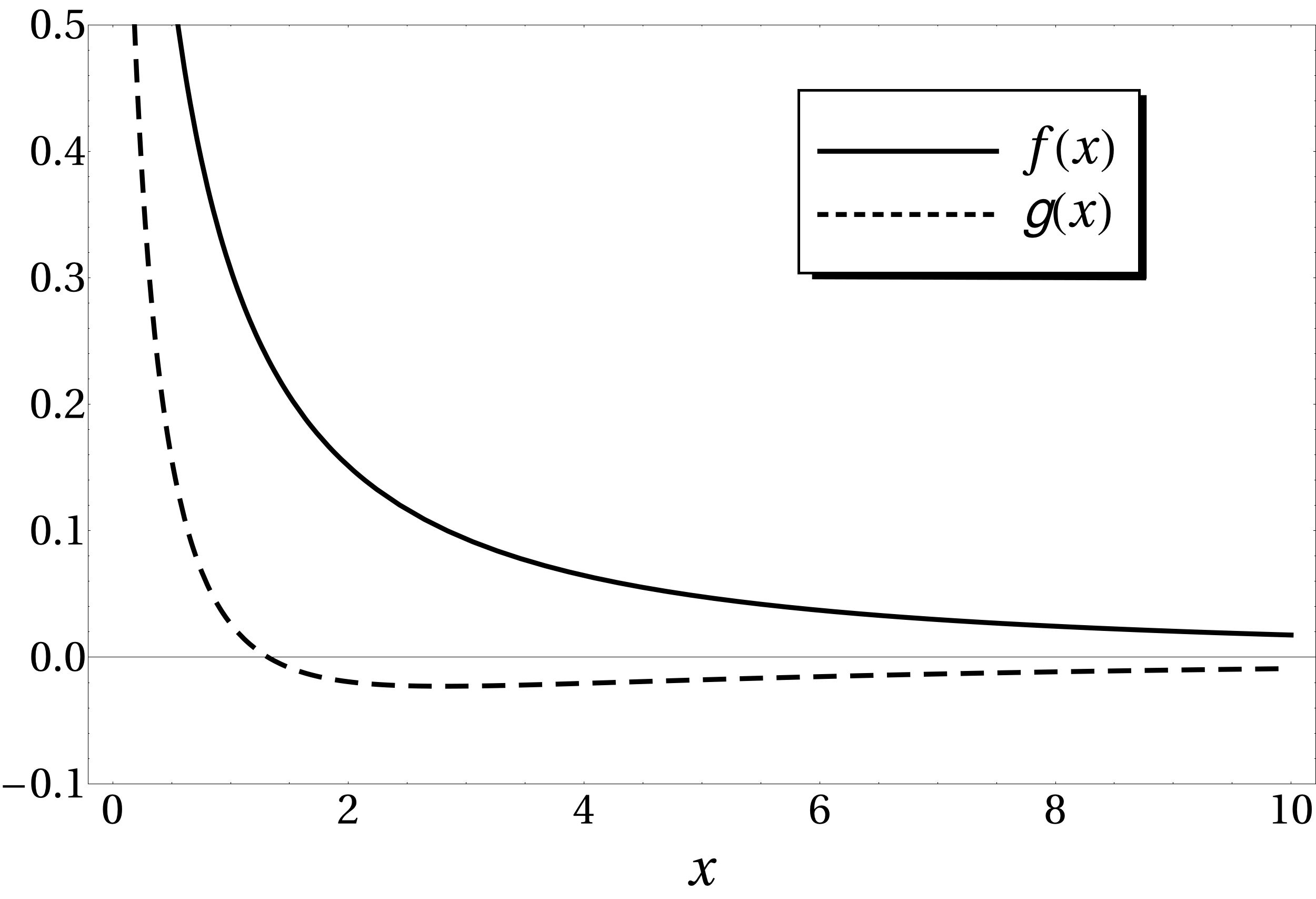}
	\caption{An horizontal line, $C>0$, determines the critical curves when crossed with the functions $f(x)$, Equation (\ref{functionf}), and $g(x)$, Equation (\ref{functiong}). Note that in $x\in(1.3182045...,\infty)$ $g(x)<0$.}
	\label{figu2}
\end{figure}

The Fig. \ref{figu2} shows the functions $f(x)$ and $g(x)$, which, as they should be, are decreasing in $x$ for $f(x)>0$ and $g(x)>0$.
The deviation angle can be calculated through the Eq. (\ref{alpha})
\begin{equation}
\alpha(x)=\frac{1}{Cx}\left(\ln\left(\frac{x}{2}\right)+\frac{2}{\left(1-x^2\right)^{1/2}}\mbox{ArcTanh}\left[\frac{(1-x)^{1/2}}{(1+x)^{1/2}}\right]\right),
\end{equation}
where the $C$ constant is given by equation (\ref{cconstant}). Now, it is straightforward that the lens equation for a mass distribution modelled by the NFW profile, reads as
\begin{eqnarray}\label{NFWlenseq}
y&=&\left|x-\frac{1}{Cx}\left(\ln\left(\frac{x}{2}\right)+\frac{2}{\left(1-x^2\right)^{1/2}}\times\right.\right.\nonumber\\
&&\qquad\left.\left.\qquad\qquad\mbox{ArcTanh}\left[\frac{(1-x)^{1/2}}{(1+x)^{1/2}}\right]\right)\right|.\qquad
\end{eqnarray}

\begin{figure}[H]
	\includegraphics[width=84mm]{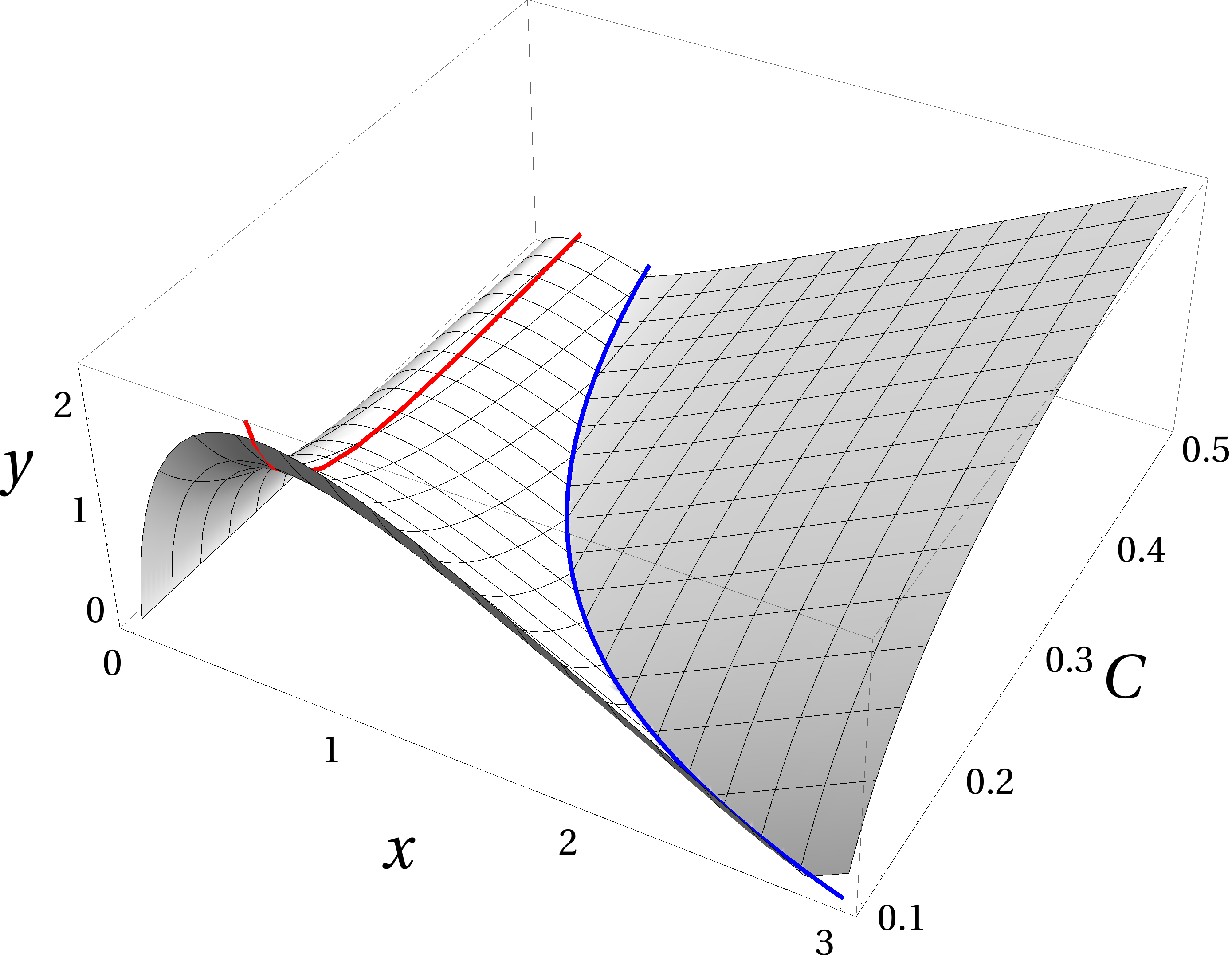}
\caption{Lens equation by a NFW model, Eq. (\ref{NFWlenseq}). As shown, the image positions will depend on the magnitud of the source and the parameters of the profile, $r_s$, $\rho_c$ and $\delta_k$, represented by $C$, Eq. (\ref{cconstant}). The local maxima and minima (red and blue lines), corresponds to the radius of the critical circles.}
\label{figu3}
\end{figure}

\begin{figure}[H]
	\includegraphics[width=84mm]{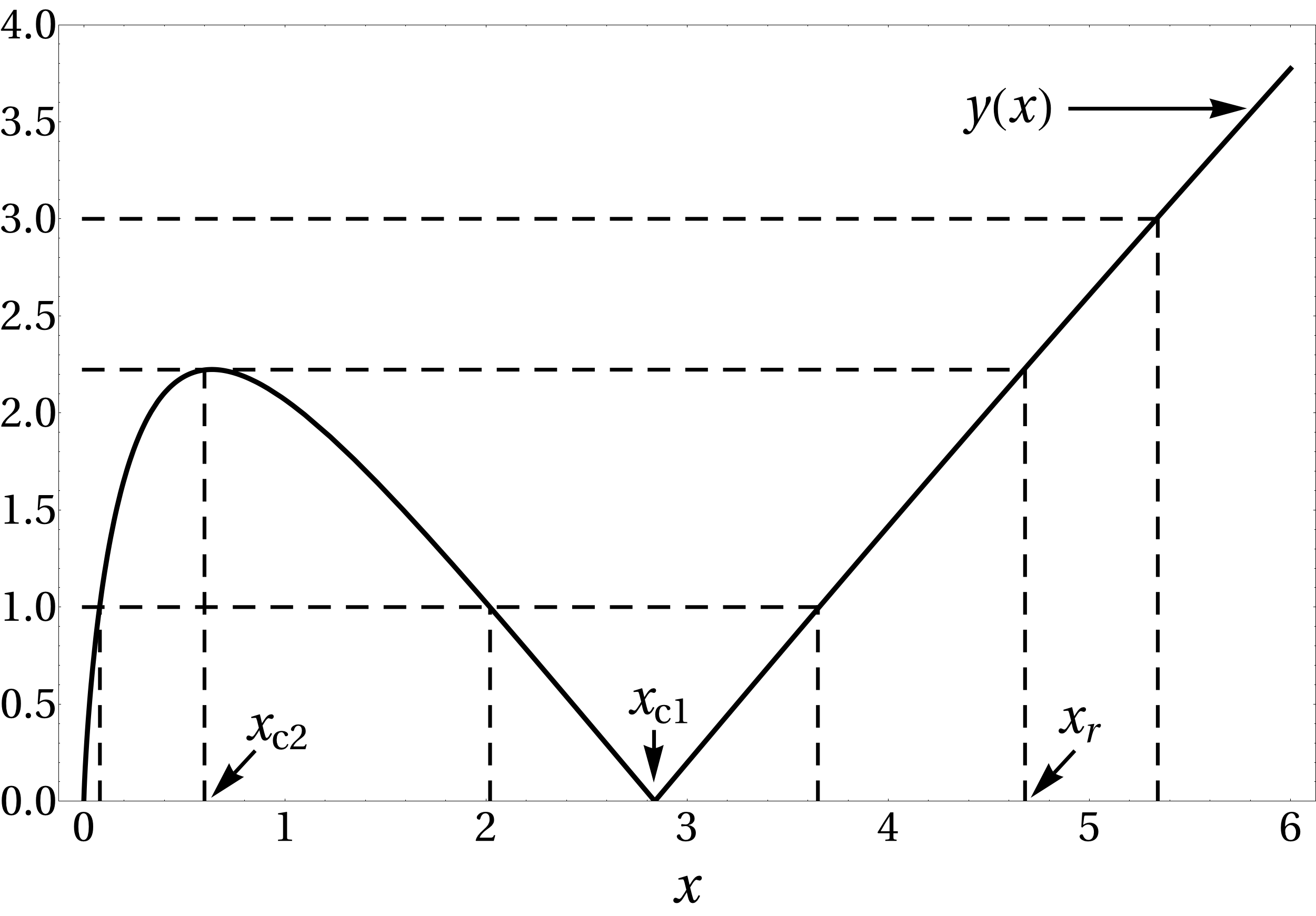}
\caption{Lens equation by a NFW model for $C=0.1$. In $x_{c1}\neq0$, the function intercepts the horizontal axis and takes its minimum value. In $x_{c2}$ the function takes its local maximum in $[0,x_{c1}]$. If $y>y(x_{c2})$ the images will be located outside the circle of radius $x_r$.}
\label{figu4}
\end{figure}

The behavior of the lens equation is shown in Fig. \ref{figu3}. There it can be seen that the local maxima and minima of the lens equation depends on the parameters of the model, these points correspond to critical curves.
The Fig. \ref{figu4} shows the lens equation in the case $C=0.1$. Depending on the source position there are four posibilities of image formation, if:
\begin{itemize}
\item $y=0$, there are infinite images (Einstein's ring of radius $x_{c1}$), if
\item $0<y<y(x_{c2})$, there are three images, the first within the circle of radius $x_{c2}$ and second one outside of it, but within of that of radius $x_{c1}$, and third outside the circle of radius $x_{c1}$, but within that of radius $x_r$, if
\item $y=y(x_{c2})$, there are two images in $x_{c2}$ and $x_r$, and if
\item $y>y(x_{c2})$, there is only one image outside of the circle of radius $x_r$.
\end{itemize}

\subsection{NFW Critical and Caustics curves}
\begin{figure}[H]
	\includegraphics[width=84mm]{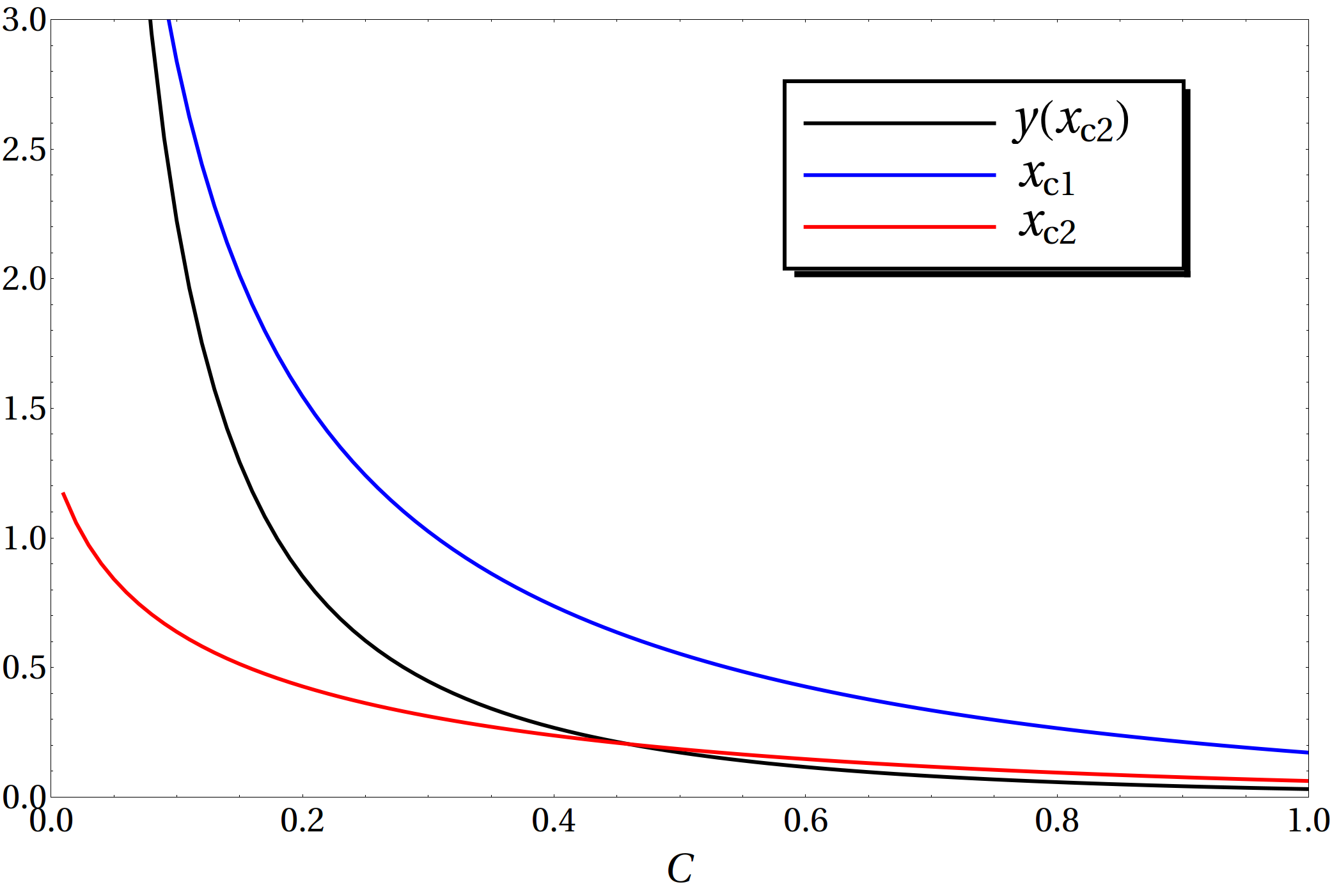}
\caption{Behavior of the points $x_{c1}$ and $x_{c2}$ where the lens equation takes its maximum values. Radius of the caustic circle associated to $x_{c2}$, as a function of $C$. This curves were found numerically.}
\label{figu5}
\end{figure}

In Fig. \ref{figu5} are displayed the local maxima of the lens equation as a function of $C$. This values are the inverse functions $f^{-1}(C)$ and $g^{-1}(C)$ and therefore, they correspond to the radii of the critical circles, in fact, their maximum values are taken when $C\rightarrow0$, where, $x_{c1}\rightarrow\infty$ and $x_{c2}\rightarrow1.3182...$. The radius of the caustic circle, also as a function of $C$, is shown in the same plot.

\subsection{NFW Shear, image positions and Magnification}
From Equation (\ref{gamma}), we find
\begin{eqnarray}\label{gammanfw}
\gamma(x)=\frac{1}{2Cx^2}\left(2\ln\left(\frac{x}{2}\right)+\frac{x^2}{1-x^2}+\frac{4-6x^2}{\left(1-x^2\right)^{3/2}}\times\right.\nonumber\\
\left.\mbox{ArcTanh}\left[\frac{(1-x)^{1/2}}{(1+x)^{1/2}}\right]\right).\quad
\end{eqnarray} 
Shear Eq. (\ref{gammanfw}) is a continuous and decreasing function over the range $(0,\infty)$, as it must be since shear is a lensing effect that should be diminish as the distance to the lens increases. In fact
\begin{equation}
\lim_{x\to0}\gamma(x)=\frac{1}{4C},
\end{equation}
\begin{equation}
\lim_{x\to1}\gamma(x)=\frac{5-3\ln 4}{6C},
\end{equation}
and
\begin{equation}
\lim_{x\to\infty}\gamma(x)=0.
\end{equation}

\begin{figure}[H]
	\includegraphics[width=84mm]{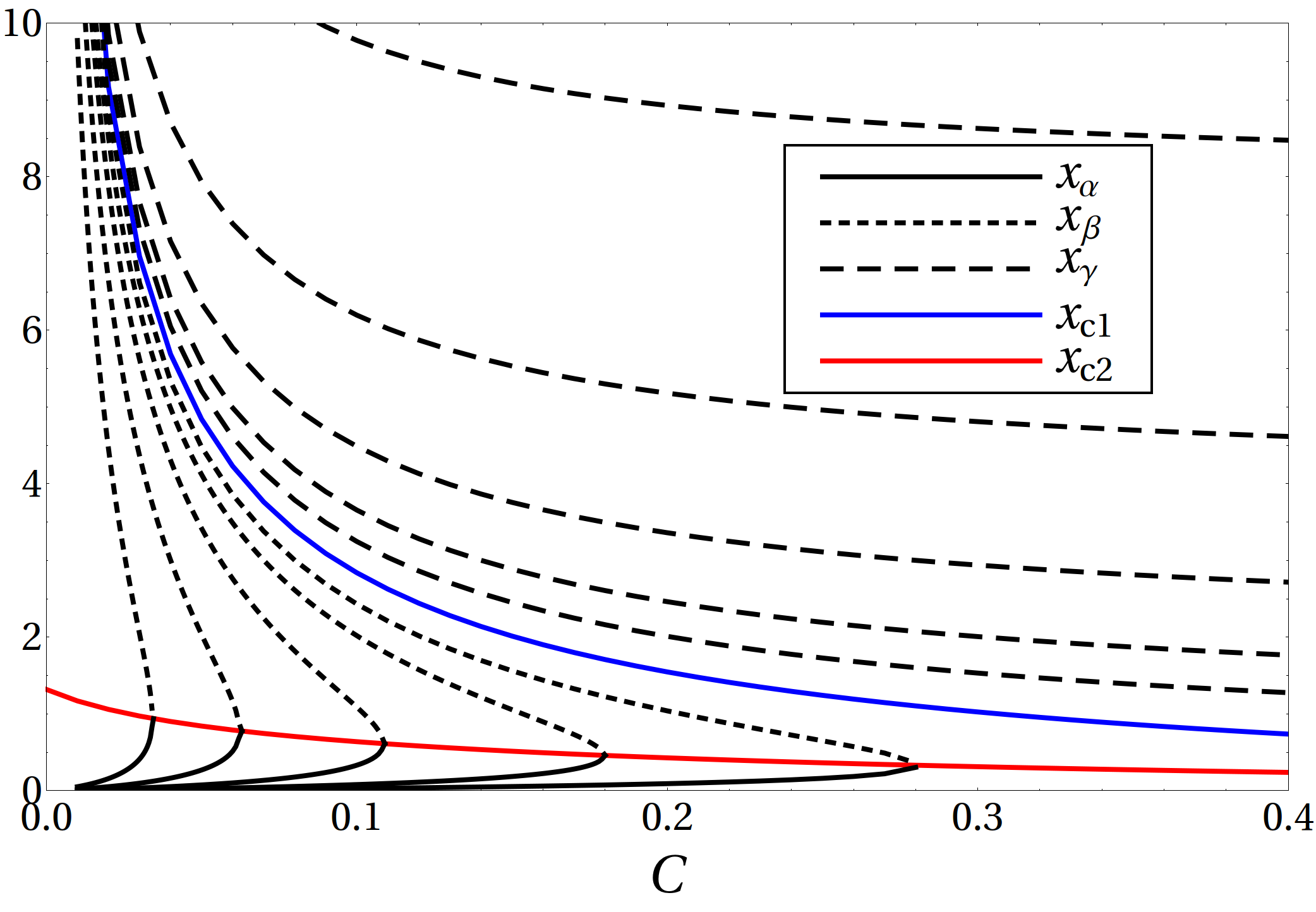}
\caption{Image positions of a point source as a function of $C$. The critical curves $x_{c1}$ (blue) and $x_{c2}$ (red) divides the source plane in three regions of image formation, that is, depending of the source position we can found up to 3 images. In agreement to Section (\ref{forimage}), solid black lines represents the position of the first image ($\alpha$), dotted lines, the second one ($\beta$), and the dashed lines, the third ($\gamma$), for each case of $y=8,4,2,1,1/2$. The three images are associated as follows: each of the curves from left to right and under $x_{c1}$ is associated with one curve from top to bottom above $x_{c1}$. If the image position approaches to zero, i.e. $y\rightarrow0$, then the Einstein Ring of radius $x_{c1}$ is formed, and the images $x_\alpha$ and $x_\beta$ go to zero, as we can see from the plot.}
\label{figu6}
\end{figure}

\begin{figure}[H]
	\includegraphics[width=84mm]{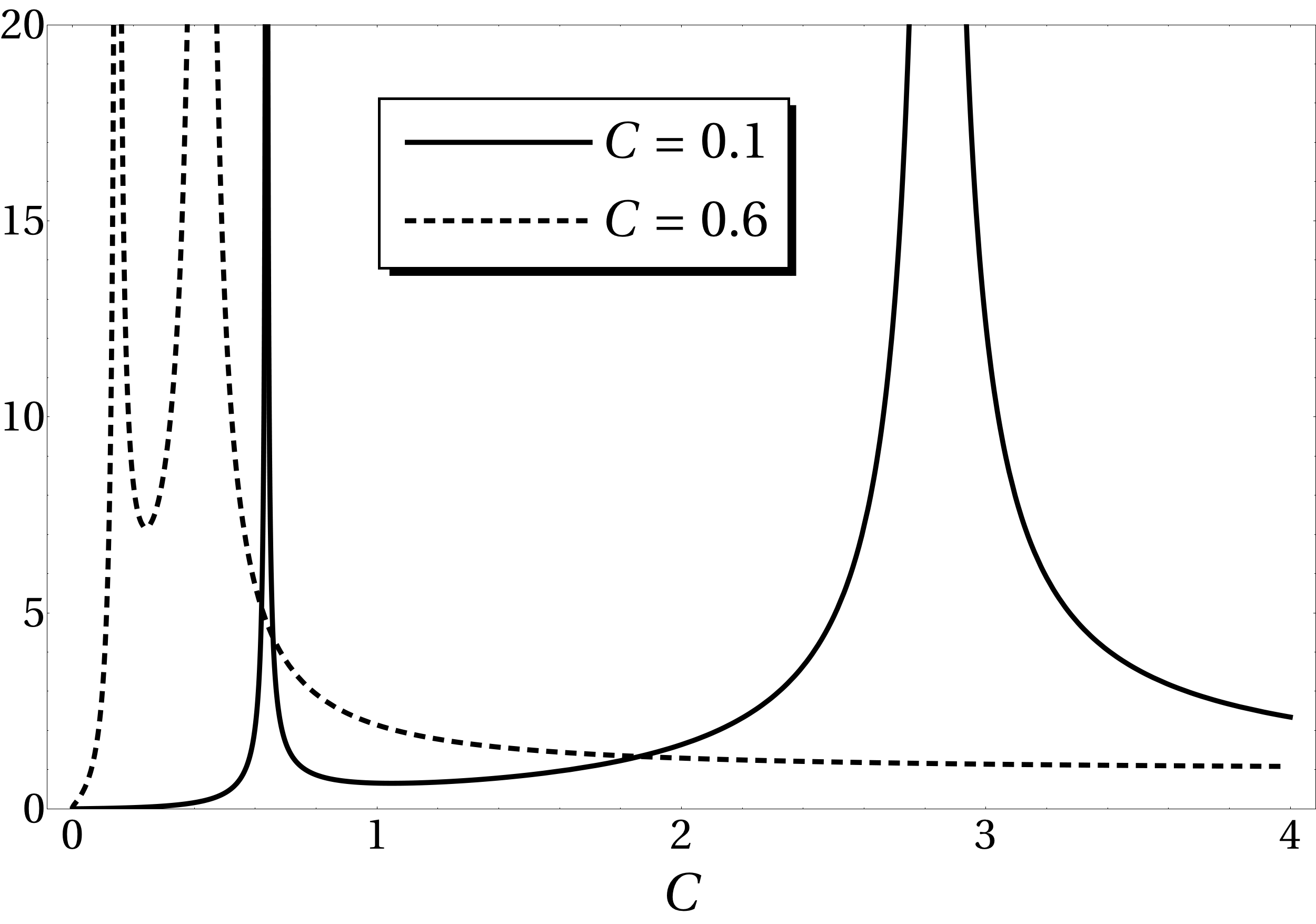}
\caption{Magnification of images in the lens plane for two values of $C$. The asymptotes will form in $x_{c1}$ and $x_{c2}$ (in each curve from right to left, respectively). Simulation of the image formation for a circular lens, magnification, critical and caustic curves generated by a lens modeled through a NFW profile is available online.}
\label{figu7}
\end{figure}

Fig. \ref{figu6} shows position of the images for different values of the source position. The change in position of the images is smaller as $C$ increases and $y$ decreases. The greater $C$ \footnote{That is, lower central density of the lens Eq. (\ref{cconstant})}, and the lower source position, the position of the $\alpha$ image tends to zero, the $\beta$ image tends to the inner critical circle and the $\gamma$ image tends to the outer critical circle.
At the same time, the magnification, given by Eq. (\ref{mu}) though Eq. (\ref{functionf}) and Eq. (\ref{functiong}) is plotted in Fig. (\ref{figu7}) for two values $C=0.1$ and $0.6$. There, we can see that the $\alpha$ image is highly demagnified when it approaches to zero, and the same occurs to the $\gamma$ image when $C$ increases.
\section{Conclusions}
In this paper we introduce a new proposal to study the gravitational lens effect by a spherically symmetric mass distribution. The main result is the use of a new function $f(x)$ which depends on the lens properties and the lens problem is described by the first order differential equation (\ref{differentialf}) which encodes all information about lensing observables. If the surface mass density of the lens is continuous, this method leads to the deflection angle in a direct way by multiplying the function for the dimmensionless coordinate $x$. We describe the critical and caustic curves through an equation that relates the function and the parameter $C$, Equation (\ref{cconstant}), of the lens which contains all the physical information of the lens and also is a function of the cosmological model.
\\
The importance of the method described in this paper is that if you resolve the equation (\ref{differentialf}) for $f(x)$, then you can find the lens observables directly in terms of that function. This implies that you do not need to solve the Poisson equation to find the deflection potential, and this is an advantage.
\\
In the case where the convergence is not a continuous function of the space, the differential equation (\ref{differentialf}) can still be used to find the $f(x)$ function, however, the deflection angle must be calculated through the Equation (\ref{alphaint}). In the Appendix \ref{appendix} we explore this approach by the point mass lens.
\\
We apply the method to a lens modelled by the NFW profile and found explicitly the function $f(x)$ in this case. The critical and caustic curves, shear, magnification and the image formation are found for this model using the formalism proposed in the first part of this paper.

\section*{Acknowledgments}

I thank Y. Villota for some helpful suggestions that improved the presentation of the paper and the Universidad Nacional de Colombia for financial support.

\appendix

\section*{Appendix A. Point Mass}\label{appendix}
Supposse a point lens at the origin of a reference frame, whose convergence, given in terms of the Dirac Delta function $\delta(x)$, is
\begin{equation}
\kappa_{PM}=\frac{M}{2\pi\Sigma_{cr}}\frac{\delta(x)}{x},
\end{equation}
With the Equation (\ref{differentialf}) we can obtain the differential equation
\begin{equation}\label{differentialpm}
x\frac{df(x)}{dx}+2f(x)-\frac{\delta(x)}{x}=0,
\end{equation}
where we take
\begin{equation}
C_{PM}=\frac{\pi\Sigma_{cr}}{M},
\end{equation}
the solution for the Equation (\ref{differentialpm}) is
\begin{equation}
f_{PM}(x)=\frac{c_1}{x^2}+\frac{\theta(x)}{x^2},
\end{equation}
where $c_1$ is the constant relate to the initial condition of the equation and $\theta(x)$ is the Heaviside function. Now, through the Equation (\ref{g}), we can obtain
\begin{equation}
g_{PM}(x)=-\frac{c_1}{x^2}+\frac{\delta(x)}{x}-\frac{\theta(x)}{x^2}.
\end{equation}
It is worth highlighting that the surface mass density of the point mass lens is not a continuos function, in fact, the Dirac Delta is a distribution, therefore, the $f_{PM}(x)$ and $g_{PM}(x)$ functions only will have meaning when we integrate them. For this, the deflection angle can not be calculated through Equation (\ref{alpha}), instead of this, we must use the Equation (\ref{alphaint})
\begin{equation}
\alpha_{PM}(x)=C_{PM}^{-1}\frac{1}{x}\int_{0}^{x}\delta(x')dx',
\end{equation}
where we can appreciate that the $c_1$ constant of the differential equation is irrelevant to the solution of the problem. So
\begin{equation}
\alpha_{PM}(x)=C_{PM}^{-1}\frac{1}{x}.
\end{equation}
This is the deflection angle found frequently in the literature which can be found from General Relativity.

\end{multicols}


\end{document}